\documentclass[a4paper,twocolumn,english,showpacs,aps]{revtex4}
\usepackage[T1]{fontenc}
\usepackage[latin9]{inputenc}
\usepackage{verbatim}
\usepackage{textcomp}
\usepackage{amsmath}
\usepackage{graphicx}

\makeatletter

\pdfpageheight\paperheight
\pdfpagewidth\paperwidth

\providecommand{\tabularnewline}{\\}

\@ifundefined{textcolor}{}
{%
 \definecolor{BLACK}{gray}{0}
 \definecolor{WHITE}{gray}{1}
 \definecolor{RED}{rgb}{1,0,0}
 \definecolor{GREEN}{rgb}{0,1,0}
 \definecolor{BLUE}{rgb}{0,0,1}
 \definecolor{CYAN}{cmyk}{1,0,0,0}
 \definecolor{MAGENTA}{cmyk}{0,1,0,0}
 \definecolor{YELLOW}{cmyk}{0,0,1,0}
}

\usepackage{epsfig}
\usepackage{dcolumn}
\usepackage{epsfig}
\usepackage{currvita}
\usepackage{times}
\usepackage{mathptm}\usepackage{upgreek}
\usepackage{babel}

\date{\today}
\usepackage{babel}

\makeatother

\usepackage{babel}
\begin{document}
\title{The electrical conductivity tensor of $\beta$-Ga$_{2}$O$_{3}$ analyzed
by van der Pauw measurements:\\
 Inherent anisotropy, off-diagonal element, and the impact of grain
boundaries }
\author{Christian Golz}
\affiliation{Department of Physics, Humboldt-Universität zu Berlin, Newton-Str.
15, D-12489 Berlin, Germany.}
\author{Zbigniew Galazka}
\affiliation{Leibniz-Institut für Kristallzüchtung, Max-Born-Str. 2, 12489 Berlin,
Germany.}
\author{Vanesa Hortelano}
\affiliation{Department of Physics, Humboldt-Universität zu Berlin, Newton-Str.
15, D-12489 Berlin, Germany.}
\author{Fariba Hatami}
\affiliation{Department of Physics, Humboldt-Universität zu Berlin, Newton-Str.
15, D-12489 Berlin, Germany.}
\author{W. Ted Masselink}
\affiliation{Department of Physics, Humboldt-Universität zu Berlin, Newton-Str.
15, D-12489 Berlin, Germany.}
\author{Oliver Bierwagen}
\affiliation{Paul-Drude-Institut für Festkörperelektronik, Leibniz-Institut im
Forschungsverbund Berlin e.V., Hausvogteiplatz 5--7, D-10117 Berlin,
Germany.}
\email{bierwagen@pdi-berlin.de}

\begin{abstract}
The semiconducting oxide $\beta$-Gallium Oxide ($\beta$-Ga$_{2}$O$_{3}$)
possesses a monoclinic unit cell whose low symmetry generally leads
to anisotropic physical properties. For example, its electrical conductivity
is generally described by a polar symmetrical tensor of second rank
consisting of four independent components. Using van der Pauw measurements
in a well-defined square geometry on differently-oriented high-quality
bulk samples and the comparison to finite element simulations we precisely
determine the ratio of all elements of the $\beta$-Ga$_{2}$O$_{3}$
3-dimensional electrical conductivity tensor. Despite the structural
anisotropy a nearly isotropic conductivity at and above room temperature
was found with the principal conductivities deviating from each other
by less than 6~\% and the off-diagonal element being $\approx3$~\%
of the diagonal ones. Analysis of the temperature dependence of the
anisotropy and mobility of differently doped samples allows us to
compare the anisotropy for dominant phonon-scattering to that for
dominant ionized-impurity scattering. For both scattering mechanisms,
the conductivites along the $a$ and $b$-direction agree within 2\,\%.
In contrast, the conductivity along $c$-direction amounts to $0.96\times$
and up to $1.12\times$ that along the $b$-direction for phonon and
ionized impurity scattering, respectively. The determined transport
anisotropies are larger than the theoretically predicted effective
mass anisotropy, suggesting slightly anisotropic scattering mechanisms.
We demonstrate that significantly higher anisotropies can be caused
by oriented extended structural defects in the form of low-angle grain
boundaries for which we determined energy barriers of multiple 10~meV. 
\end{abstract}
\maketitle

\section{Introduction}

$\text{\ensuremath{\beta}-Ga}_{\text{2}}\text{O}_{\text{3}}$, the
thermodynamically stable polymporph of solid $\text{Ga}_{\text{2}}\text{O}_{\text{3}}$,
is a promising material for several applications such as high power
electronics\cite{Higashiwaki,HigashiwakiJessen} and deep UV photo
detectors\cite{Oshima,HigashiwakiJessen}, and it can be used for
high-temperature gas sensors\cite{Lampe}. $\text{\ensuremath{\beta}-Ga}_{\text{2}}\text{O}_{\text{3}}$
has a monoclinic lattice structure, which corresponds to the C2/m-space
group with lattice parameters of $a=$12.23~Å, $b=$3.04~Å, and
$c=$5.80~Å and an angle of $103.7^{\circ}$ between $a$ and $c$-axis\cite{Geller}.
Due to this angle, the basis vectors of the unit cell, $a$ and $c$,
are not orthogonal to the (100) and (001) planes in $\text{\ensuremath{\beta}-Ga}_{\text{2}}\text{O}_{\text{3}}$.
The low symmetry of the unit cell is prone to result in anisotropic
physical properties.

Notable anisotropies in $\text{\ensuremath{\beta}-Ga}_{\text{2}}\text{O}_{\text{3}}$
have been found in the thermal conductivity \cite{Guo,Santia,Handwerg}
and the dielectric function by the polarization-dependent refractive
index \cite{Sturm} and fundamental onset of optical absorption \cite{Ueda,Onuma,Sturm}.
In addition, slight anisotropies of the high-frequency and static
dielectric constant of $\text{\ensuremath{\beta}-Ga}_{\text{2}}\text{O}_{\text{3}}$
have been reported by theory and experiment\cite{He,Liu,Schubert,Fiedler2019,Ghosh2016}.
\begin{table*}
\caption{Overview of the literature on transport anisotropies and associated
methods in $\text{\ensuremath{\beta}-Ga}_{\text{2}}\text{O}_{\text{3}}$.
The anisotropy ``Aniso.'' is the ``Quantity'' in direction ``Dir.1''
divided by that in ``Dir.2'' at room-temperature electron concentration
``$n$@RT''. Directions $a,\,b,\,c$ are parallel to the $a,\,b,\,c$
axes of the unit cell, and and $a${*}$,\,c${*} are parallel to the
(100), (001) surface normals. ``n\textbackslash s'' denotes ``not
specified''. \#The samples used for different transport directions
had different electron concentrations. {*}The anisotropy of less than
1.1 for $\sigma$ (line 3) was calculated from a van der Pauw resistance
anisotropy ratio of less than 1.3 given in Ref.~\cite{Irmscher}
using the conversion into conductivity anisotropy elaborated in Ref.~\cite{Bierwagen}.}
\begin{tabular}{llccccc}
\hline 
Quantity  & Method  & Dir.1  & Dir.2  & Aniso.  & $n$(cm$^{-3}$) @RT & Ref.\tabularnewline
\hline 
$\sigma$  & 4-probe  & $b$  & $c$  & 17 (extrinsic?)  & $5.2\times10^{18}$ & \cite{Ueda} \tabularnewline
$\sigma$  & van der Pauw  & $b$  & $c$  & 2 (twins)  & $>10^{18}$%
{} & \cite{Fiedler}\tabularnewline
$\sigma$  & van der Pauw  & all  & all  & <1.1 {*}  & $5\times10^{16}-5\times10^{17}$ & \cite{Irmscher}\tabularnewline
\hline 
$\mu$  & Hall bar  & $c$  & $a$, $b$  & 1.2 \# & $4\times10^{17}$, $7\times10^{17}$, $9\times10^{17}$ & \cite{Villora}\tabularnewline
$\mu$  & MOSFET channel  & $a*$  & $c$  & 1.1  & $3\times10^{17}$ & \cite{Wong}\tabularnewline
$\mu$  & ellipsometry  & $a$, $b$  & $c*$  & 2  & $3.5\times10^{18}$ & \cite{Schubert}\tabularnewline
$\mu$  & optical Hall  & all  & all  & <1.1  & $4\times10^{18}$, $6\times10^{18}$ & \cite{Knight}\tabularnewline
$\mu_{PLOPS}$  & theory  & $c*$  & $a$  & 1.18 to 1.40  & $10^{17}$ to $10^{20}$ & \cite{Kang}\tabularnewline
$\mu_{PLOPS}$  & theory  & $a$  & $b$  & 0.85 to 0.78  & $10^{17}$ to $10^{20}$ & \cite{Kang}\tabularnewline
$\mu_{PLOPS}$  & theory  & $c*$  & $b$  & 1.01 to 1.09  & $10^{17}$ to $10^{20}$ & \cite{Kang}\tabularnewline
$\mu_{IIS}$  & theory  & n\textbackslash s  & n\textbackslash s  & estimated <1.4  & - & \cite{Kang}\tabularnewline
$\mu_{PLOPS}$  & theory  & $c*$  & $b$  & 0.64 to 0.92  & $5\times10^{17}$ to $10^{19}$ & \cite{Ghosh2017}\tabularnewline
$\mu_{IIS}$  & theory & any & any & assumed 1 & $5\times10^{17}$ to $10^{19}$ & \cite{Ghosh2017}\tabularnewline
$\mu_{PLOPS+IIS}$  & theory & $c*$  & $b$  & 0.69 to 0.92  & $5\times10^{17}$ to $10^{19}$ & \cite{Ghosh2017}\tabularnewline
\hline 
$m^{*}$  & optical Hall  & all  & all  & <1.1  & $4\times10^{18}$, $6\times10^{18}$ & \cite{Knight}\tabularnewline
$m^{*}$  & theory  & all  & all  & <1.05  & - & \cite{Furthmuller,He,Peelaers,Yamaguchi}\tabularnewline
\end{tabular}\label{tablit} 
\end{table*}
The conductivity anisotropy is given by the ratio of the electrical
conductivity in two different defined directions. The same holds true
for the mobility tensor $\bar{\mu}$ as $\bar{\sigma}=en\bar{\mu}$,
as the charge carrier density $n$ and electronic charge $e$ are
scalars.

For electronic device applications such as transistors, a mobility
anisotropy would translate into an increased performance of the devices
for a certain orientation compared to the crystallographic directions.
As shown in Tab.~\ref{tablit}, quite contradictory experimental
values on the conductivity anisotropy can be found in literature,
whereas no estimate has been reported so far on the off-diagonal element
of conductivity. Reported conductivity anisotropies are ranging from
17 times higher conductivity for the $b$-direction compared to the
$c$ direction \cite{Ueda}, over the same mobility in $a$ and $b$
direction and 1.2 times higher conductivity in $c$-direction \cite{Villora}
to a negligible anisotropy \cite{Irmscher}, where no direction was
favored for transport. In metal-oxide field effect transistors (MOSFETs),
a 10\% larger channel mobility in the $a${*} direction compared to
the $c$ direction has been observed\cite{Wong}. Using ellipsometry,
the same mobility in $a$ and $b$, but only half the mobility in
$c$-direction has been measured \cite{Schubert}. In combination
with an external magnetic field, the same method, termed optical Hall
effect, has yielded rather isotropic mobilities (within $\approx10\,\%$)\cite{Knight}.

As the mobility depends on the effective mass $m^{*}$ and scattering
time $\tau$, $\mu=\frac{e\tau}{m^{*}}$, its anisotropy is determined
by the anisotropy of $m^{*}$ and $\tau$. Surprisingly, first-principles
calculations from several groups using different methods arrive at
a fairly isotropic effective electron mass \cite{Furthmuller,He,Peelaers,Yamaguchi}
with anisotropies typically below 1.05, which is confirmed by experimental
data of the effective mass using optical Hall effect measurements\cite{Knight}.

The scattering time is related to the dominant scattering mechanism,
polar longitudinal optical-phonon scattering (PLOPS) and ionized impurity
scattering (IIS).\cite{Ma,Kang,Ghosh2017} At and above room temperature,
the mobility is limited by PLOPS\cite{Ma,Kang}, whereas IIS is dominating
at high impurity- or point-defect densities (including compensating
impurities/point defects) or lower temperatures. Depending on the
scattering mechanism the monoclinic symmetry of $\text{\ensuremath{\beta}-Ga}_{\text{2}}\text{O}_{\text{3}}$
suggests more or less anisotropic scattering times. Hence, the change
of dominant scattering mechanism with temperature likely leads to
a temperature-dependent anisotropy of the scattering rates. Recent
first-principles calculations of the electron mobility limit of $\text{\ensuremath{\beta}-Ga}_{\text{2}}\text{O}_{\text{3}}$,
indeed, suggested a moderate anisotropy with higher mobility in $c*$-
than in $a$- or $b$-direction.\cite{Kang} In that work, the PLOPS-limited
mobility (neglecting IIS) in the $c*$-direction has been predicted
to be up to 9\,\% and 40\,\% higher than in the $a$- and $b$-
direction, respectively, to become more isotropic with decreasing
electron concentrations, and anisotropies of up to 1.4 for the IIS-limited
mobility were estimated. Another first principles study,\cite{Ghosh2017}
predicts for the PLOPS-limited mobility opposite trends with a 36\,\%
lower mobility in the $c*$- than in the $b$-direction, which becomes
more isotropic with increasing electron concentration, and isotropic
IIS.

Besides the intrinsic material properties, samples might be extrinsically
anisotropic on the average due to oriented extended defects like grain-
and twin boundaries. As an example, Ref.~\cite{Fiedler} found a
room-temperature transport anisotropy as high as 2 in Ga$_{2}$O$_{3}$(100)
thin films containing a high density of incoherent twin boundaries.
Likewise, the high anisotropy of 17 from Ref.~\cite{Ueda} has been
interpreted in terms of extrinsic causes by Refs.~\cite{Irmscher,Kang}.

In this paper we experimentally determine the intrinsic anisotropy
and relative magnitude of the off-diagonal element of the conductivity
tensor of $\text{\ensuremath{\beta}-Ga}_{\text{2}}\text{O}_{\text{3}}$
with high accuracy (uncertainty of 2\,\%) to shed light on the conflicting
values of published theoretical and experimental transport anisotropy.
The extracted conductivity is surprisingly isotropic and we find large
transport anisotropies to be extrinsically caused by extended defects.

\section{Samples and Method}

\subsection{Samples}

To largely rule-out the extrinsic effect of extended defects on transport
anisotropy, semiconducting bulk substrates were chosen as sample material
with highest structural quality available. Square shaped $5\times5$~mm$^{2}$
(0.5~mm to 0.7~mm thickness) wafers with different orientations,
i. e. Czochralski-grown \cite{Galazka} (100) and (001) and edge-defined
film fed grown \cite{Kuramata} ($\bar{2}$01) (from Tamura Corporation)
were investigated in this work. The edges of the squares were oriented
along low-index crystallographic directions. The samples are described
in Tab.~\ref{tabxrd}. The two (1~0~0) and (0~0~1) oriented samples
G100a and G001a were prepared from the same boule. To study the impact
of extended defects on the transport anisotropy a (1~0~0)-oriented
sample containing low angle grain boundaries, G100c, was prepared.

Disk-shaped ohmic contacts of 100~$\upmu$m to 450~$\upmu$m diameter
close to the corners of the sample were reproducibly defined on the
top surface of the samples using shadow masks and photolithography
to minimize geometry errors. A distance of the contacts from the sample
edges of 700~$\upmu$m to 850~$\upmu$m was chosen to prevent unintentional
contacting the side of the sample by individual contacts. This approach
ensures reproducible and well-defined current injection (only through
the top surface) by all four contacts to provide a current distribution
that can be readily compared to the results of our corresponding finite-element
calculations. All contacts were deposited by electron beam evaporation
of Ti/Pt/Au (20~nm/20~nm/150~nm), followed by rapid thermal annealing
at $480^{\circ}$C for 60~s in N$_{2}$. The Pt layer served as diffusion
barrier as deterioration of Ti/Au contacts was observed upon annealing.

\begin{table*}[t]
\caption{Overview of the results of the X-Ray diffraction measurements and
the properties of the samples: orientation, doping, charge carrier
density, and assessed quality. The surface orientation is given by
the numbers in the sample name. UID refers to unintentionally doped
samples.}
\begin{tabular}{lcclcccc}
\hline 
Sample  & Orientation; Edges  & Doping  & Electron  & Quality  & XRD $\omega$  & $\omega$ rocking  & %
\tabularnewline
name  &  &  & density{[}$cm^{-3}${]}  &  & reflection  & curve fwhm ($^{\circ}$)  & %
\tabularnewline
G100a  & (1 0 0); {[}0 1 0{]}, {[}0 0 1{]}  & UID  & $2.5\times10^{17}$ & high & (4 0 0) & 0.012 & %
\tabularnewline
G100b  & (1 0 0); {[}0 1 0{]}, {[}0 0 1{]}  & UID  & $8.1\times10^{16}$  & high  & (4 0 0)  & 0.041  & %
\tabularnewline
G100c  & (1 0 0); {[}0 1 0{]}, {[}0 0 1{]}  & UID  & $5.9\times10^{17}$  & low (extended defects)  & (4 0 0)  & 0.180  & %
\tabularnewline
G001a  & (0 0 1); {[}1 0 0{]}, {[}0 1 0{]}  & UID  & $4.3\times10^{17}$  & high  & (0 0 2)  & 0.015  & %
\tabularnewline
G-201a  & ($\bar{2}$ 0 1); {[}1 0 2{]}, {[}0 1 0{]}  & Sn  & $5.9\times10^{18}$  & high  & ($\bar{2}$ 0 1)  & 0.034  & %
\tabularnewline
\hline 
\end{tabular}\label{tabxrd} 
\end{table*}
The electron densities of the samples were in the range of $8\times10^{16}~$cm$^{-3}$
to $6\times10^{18}~$cm$^{-3}$at room temperature.

\subsection{Structural characterization}

\begin{figure}
\includegraphics[width=4.2cm]{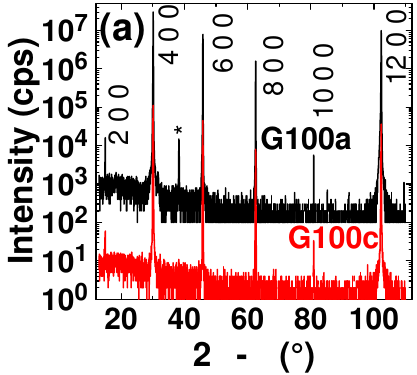}\includegraphics[width=4.2cm]{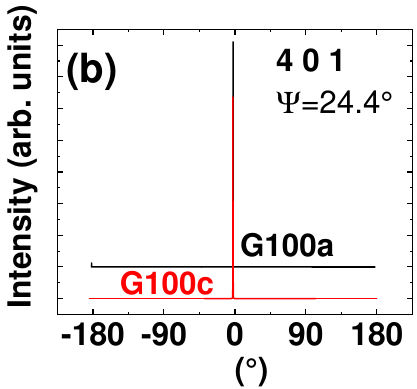}
\caption{\label{g2to}XRD scans to determine the substrate orientation shown
exemplarily for samples G100a and G100c. (a) Symmetric on-axis $2\Theta-\omega$
scans showing the presence of reflexes, labeled by their Miller indices,
only related to the 1~0~0 wafer orientation. The peak marked by
``{*}'' is related to Au(1~1~1) of the ohmic contact. (b) $\phi$-scan
of the 4~0~1 reflex. The one-fold rotational symmetry indicates
single crystalline material without rotational domains or twins.}
\end{figure}

The crystal quality of the wafers was assessed by X-ray diffraction
(XRD) using Cu-K$\alpha$ radiation and a 1~mm detector slit. Wide-range,
symmetric, on-axis, $2\Theta-\omega$ scans confirmed phase-pure material
by the presence of only the $\text{\ensuremath{\beta}-Ga}_{\text{2}}\text{O}_{\text{3}}$
reflexes belonging to the specified wafer orientation. Figure~\ref{g2to}(a)
shows an example for the two (100)-oriented wafers G100a and G100c.
To detect the potential existence of twins or rotational domains,
off-axis XRD peaks were measured by $\Phi$-scans with rotational
angle $\Phi$ around the surface normal. In these scans, the off-axis
diffraction peaks were measured in skew-symmetric geometry with the
sample tilted by the angle $\Psi$. The presence of a single peak
in the $\Phi$-scan of the 4~0~1 reflex for G100a, b, and c as well
as G001a, exemplarily shown in Fig.~\ref{g2to}(b) for G100a and
G100c, confirms the absence of twins or rotational domains in the
(1~0~0)- and (0~0~1) oriented samples. Likewise, a one-fold rotational
symmetry was confirmed for the (-2~0~1) oriented sample G-201a by
the presence of only one peak in the $\Phi$-scan of the 4~0~0 reflex
(not shown).

\begin{figure}
\includegraphics[width=4cm]{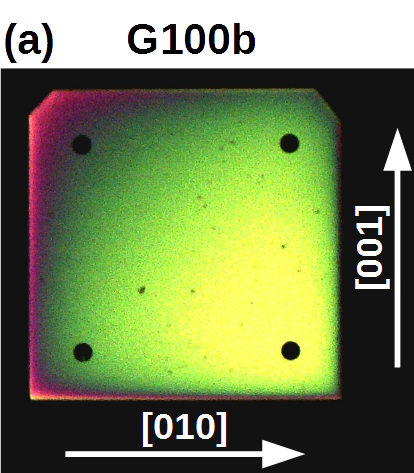}\hspace*{5mm}\includegraphics[width=4cm]{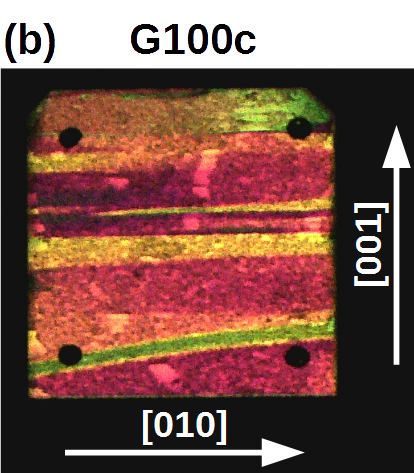}

\qquad{}\includegraphics[width=4.2cm]{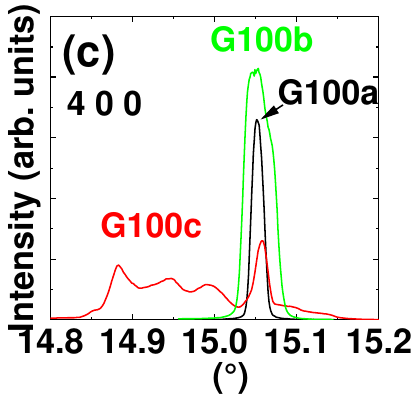}\includegraphics[width=4.2cm]{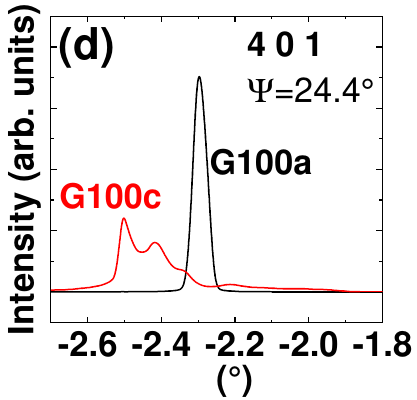}\caption{\label{go3}Comparison of (100)-oriented wafers with different structural
quality: G100a, b without- and G100c with low angle grain boundaries.
(a), (b) Photographs of G100b (a) and G100c (b) placed between two
crossed-polarizers with white light illumination from the backside.
The crystallographic directions are indicated by white arrows. Horizontal
lines in the sample G100c indicate the grain boundaries. No such defects
are visible in the sample G100b. The dark dots are the disc-shaped
ohmic contacts. (c) XRD $\omega$-rocking curves of the 4~0~0 reflex
of samples G100a, b and c. (d) XRD detailed $\Phi$-scan of the 4~0~1
reflex of G100a and G100c.}
\end{figure}

To assess the crystal quality in more detail, $\omega$-rocking curves
of the on-axis substrate reflections, sensitive to lattice tilting,
were taken for all samples. The low full width at half maximum of
these curves below 0.1~°, documented in Tab.~\ref{tabxrd}, confirms
the comparably high crystalline quality. The only exception is, sample
G100c with a rather broad rocking curve. Figure~\ref{go3} compares
this sample to the other two (1~0~0) oriented ones. Photographs
of the wafers placed between two crossed polarizers with white light
illumination from the backside exhibit a comparably homogeneous contrast
for G100b shown in Fig.~\ref{go3}(a) but an inhomogeneous contrast
with stripes oriented approximately along the {[}0~1~0{]} direction
for G100c shown in Fig.~\ref{go3}(b). These stripes represent single
crystalline domains that are slightly twisted and tilted with respect
to each other as indicated by the multi-peak structure of the rocking
curve (Fig.~\ref{go3}(c)) and detailed $\Phi$-scan (Fig.~\ref{go3}(d)),
respectively. The inhomogeneous contrast is due to the trichroism
of $\beta$-Ga$_{2}$O$_{3}$ resulting in differently colored regions
for differently oriented grains.

Hence, all samples can be considered single crystalline except for
G100c, which consists of a low number of single crystalline domains
with low angle (<0.3°) grain boundaries. This sample allows us to
investigate the influence of low-angle grain boundaries on the transport
properties. 

\subsection{Extraction a 2-dimensional conductivity anisotropy by van der Pauw
measurements and simulation}

In order to investigate the effect of the scattering mechanisms and
energy barriers due to low-angle grain boundaries on the conductivity
anisotropy, transport measurements were conducted. These measurements
were done in van der Pauw geometry (square with four contacts close
to the corners) at temperatures between 10~K and 375~K in a closed-cycle
helium refrigerator.

Hall measurements were performed in a magnetic field of $B=\pm0.5$~T
oriented perpendicular to the substrate surface. For Hall measurements,
two contacts diagonal to each other are used for the current $I$
respectively for measuring the voltage $V$. Using the two contacts
on one edge parallel $x$ of the square sample for applying $I$ and
those on the opposite edge for measuring $V$, the four-terminal resistance
along $x$, $R_{x}$, is determined. The same is done for the other
two edges (perpendicular to the former ones), which yields $R_{y}$.
These $R_{x}$ and $R_{y}$ are used as described in the original
work by van der Pauw to calculate the geometrical average of the anisotropic
sheet resistance $R_{ave}=1/\sqrt{\sigma_{x}\sigma_{y}}$. Their ratio
$A_{vdP}=R_{y}/R_{x}$, can be translated into the conductivity ratio
$A=\sigma_{x}/\sigma_{y}$ along these axes using finite element methods
(FEM) simulations of the current- and voltage distribution for each
sample.

We recently described in detail how we used this method for the determination
of the transport anisotropy of another semiconducting oxide with anisotropic
crystal structure, SnO$_{2}$,\cite{BierwagenGalazka} and compared
the method to the Hall bar geometry \cite{Bierwagen}.

\begin{figure}
\includegraphics[scale=0.3]{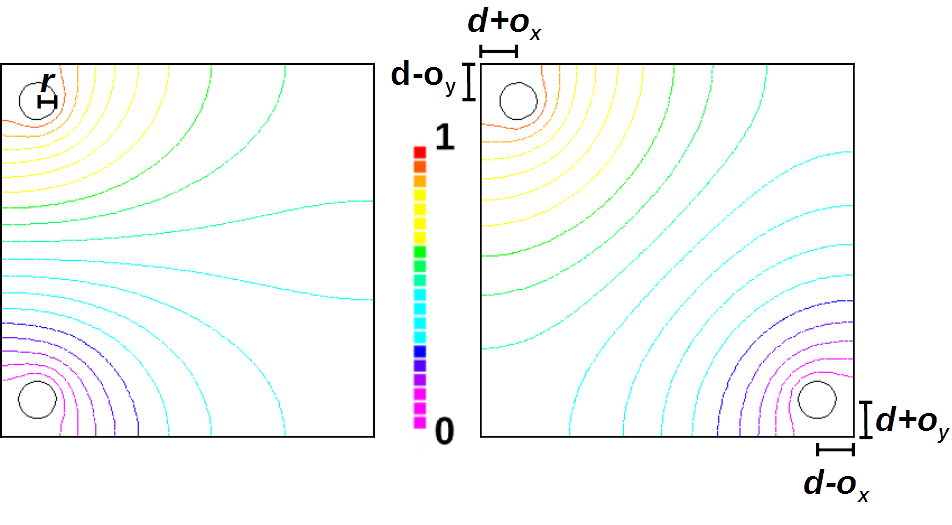} \caption{\label{simx2}Potential distribution calculated by finite element
simulation of a square shaped van der Pauw sample in the configuration
used for resistivity anisotropy measurements (left) and Hall-effect
measurements (right). $r$ denotes contact radius, $d$ is the distance
from the contact to the edge, and $o_{x}$ and $o_{y}$ are the offset
of the contact position in these directions. In this image, $r,d,o_{x}$,
and $o_{y}$ are 0.05, 0.1, 0, and 0 times the sample length, anisotropy
$A=1$ and $B=0$.}
\end{figure}

For each sample, geometry details were derived from micrograph images,
including the size, position, and shape of the contacts. Using our
FEM simulations, the impact of deviations from the ideal van der Pauw
geometry for each sample (such as extended size- and position of the
contacts away from the sample edges) on measured sheet resistance,
its anisotropy, as well as electron concentration $n$ and resulting
average mobility $\mu_{ave}=\sqrt{\mu_{x}\mu_{y}}$ are correctly
accounted for. Examples of a potential distribution resulting from
the FEM simulation can be found in Fig.~\ref{simx2}. The relation
$A(A_{vdP})$ and correction factors for $R_{ave},\,n,\,\mu_{ave}$
derived from FEM simulations for the geometries used in this work
is summarized in Tabs.~\ref{tabfem} and \ref{tabfem2}, respectively,
as well as Ref.~\cite{BierwagenGalazka}.

\begin{table}[h]
\caption{Relation between conductivity anisotropy $A$ and van der Pauw resistance
anisotropy $A_{vdP}$ calculated by two-dimensional FEM simulations
for square samples with different sizes and positions (offset $o_{y}$
as shown in Fig.~\ref{simx2}) of the disk-shaped contacts. All sizes
are relative to the edge length of the square sample.\label{tab:AvsAvdP}}
\begin{tabular}{lccccr}
\hline 
Contact  & Distance contact  & Offset of  & $A$  & $A_{vdP}$  & \tabularnewline
radius  & to the edge  & contact position $o_{y}$  &  &  & \tabularnewline
\hline 
all values  & all values  & 0  & 1  & 1  & \tabularnewline
0.05  & 0.2  & 0.1  & 1  & 1.14  & \tabularnewline
0.01  & 0.01  & 0  & 1.1  & 1.35  & \tabularnewline
0.05  & 0.1  & 0  & 1.1  & 1.32  & \tabularnewline
0.05  & 0.2  & 0  & 1.1  & 1.26  & \tabularnewline
0.01  & 0.01  & 0  & 2  & 9.4  & \tabularnewline
0.05  & 0.1  & 0  & 2  & 7.7  & \tabularnewline
0.05  & 0.2  & 0  & 2  & 5.4  & \tabularnewline
\hline 
\end{tabular}\label{tabfem} 
\end{table}
\begin{table}[h]
\caption{Correction factors for quantities derived from van der Pauw and Hall
measurements for square samples with different sizes and positions
(offset $o_{y}$ as shown in Fig.~\ref{simx2}) of the disk-shaped
contacts, calculated using two-dimensional FEM simulations. Measured
quantities are multiplied by the correction $F$ to obtain the ``true''
quantities. All sizes are relative to the edge length of the square
sample.}
\begin{tabular}{lcccccr}
\hline 
Contact  & Distance contacts  & Offset of  & $F$ for  & $F$ for  & $F$ for  & \tabularnewline
radius  & to the edge  & contact position $o_{y}$  & $R_{ave}$  & $n$  & $\mu_{ave}$  & \tabularnewline
\hline 
0  & 0  & 0  & 1  & 1  & 1  & \tabularnewline
0.05  & 0.1  & 0  & 1.01  & 0.84  & 1.18  & \tabularnewline
0.05  & 0.2  & 0  & 1.10  & 0.58  & 1.57  & \tabularnewline
0.01 & 0.15 & 0 & 1.02 & 0.80 & 1.23 & \tabularnewline
0.05  & 0.2  & 0.1  & 1.09  & 0.60  & 1.53  & \tabularnewline
\hline 
\end{tabular}\label{tabfem2} %
\end{table}
Obtaining reliable values of $A$ crucially depends on a precisely
defined contact geometry as any deviation from a square arrangement
by an aspect ratio $L_{x}/L_{y}$ of a sample with a length of $L_{x}$
and $L_{y}$ impacts the resulting conductivity (or mobility-) anisotropy
$A$ by a factor $(L_{x}/L_{y})^{2}$\cite{Bierwagen}. We tested
the accuracy of our experimental technique using several isotropic
Si bulk samples, a highly perfect semiconductor, and contact geometry
defined by photolithography. The resulting deviations of $A$ from
unity was $<0.7\,\%$ for each of these samples, which can be seen
as the intrinsic geometrical uncertainty of our experiment. A potential
experimental artifact that can impact the observed in-plane conductivity
anisotropy is the microscopic location of the current injection into
the semiconductor under the ohmic contact pad. In our simulations
we are assuming current injection along the periphery of the contact.
Low electron concentrations and low temperatures can, however, lead
to a significant increase of contact resistance and inhomogeneous
current injection underneath the contacts (contact freeze-out). For
the lithographically defined contacts with diameter of 300(100)~\textmu m
whose center has a distance of 850(750)~\textmu m to nearest sample
edges a maximum anisotropy artifact can arise for current injection
only at one point on the contact pad that is located e.g. in the horizontal
direction towards the outer edge of the sample but in the vertical
direction towards the center of the sample. In this extreme case,
the maximum apparent van der Pauw anisotropy for isotropic conductivity
would be 1.33(1.09), which corresponds to an error for the extracted
conductivity anisotropy $\pm11$\,\%($\pm3$\,\%). Experimental
comparison of the extracted transport anisotropy of sample G100a and
and G100b with two different contact sets (300\,\textmu m and 100\,\textmu m
diameter), each, yielded an agreement within 1--2\% between the different
contact sets, indicating negligible inhomogeneity of current injection
underneath the contacts.

\subsection{The 3-dimensional conductivity tensor}

\begin{figure}
\includegraphics[scale=0.4]{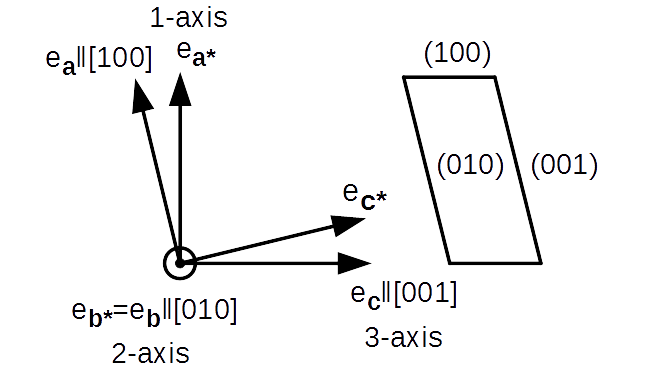} \caption{\label{cell}The unit cell of $\text{\ensuremath{\beta}-Ga}_{\text{2}}\text{O}_{\text{3}}$projected
along the $b$-axis (= {[}010{]} direction). The unit basis vectors
along the directions $a,\,b,\,c$ and the unit vectors perpendicular
to the (100), (010), (001) planes (along $a${*}$,\,b${*}, and $c${*})
are shown. The 1-axis=$e_{a}${*}, 2-axis=$e_{b}$, and 3-axis=$e_{c}$
refer to the chosen Cartesian coordinate system to describe the conductivity
tensor in this work ($a${*}$bc$-system). The angle between $a$
and $a${*} as well as $c$ and $c${*} is $13.7^{\circ}$. The vectors
$a,\,a${*}$,\,c,\,c${*} are in the shown plane and perpendicular
to $b=b${*}. }
\end{figure}
Fig.~\ref{cell} schematically shows the monoclinic unit cell of
$\text{\ensuremath{\beta}-Ga}_{\text{2}}\text{O}_{\text{3}}$ along
with the unit basis vectors along $a,\,b,\,c$ as well as unit vectors
perpendicular to the lowest index planes along $a${*}$,\,b${*}$,\,c${*}.
The vector representation of physical properties requires an orthonormal
system. Two different systems $abc${*} and $a${*}$bc$ are chosen
by different authors. We will use the $a${*}$bc$-system by aligning
the 1-axis along (100) surface normal $a${*}, the 2-axis along $b$
and the 3-axis along the $c$-axis as indicated in Fig.~\ref{cell}.
In this system, the conductivity tensor $\bar{\sigma}$ of $\text{\ensuremath{\beta}-Ga}_{\text{2}}\text{O}_{\text{3}}$,
that relates electric field $\vec{E}$ and current density $j$ by
$\vec{j}=\bar{\sigma}\vec{E}$, can be expressed at zero magnetic
field as:\cite{Parisini} 
\begin{equation}
\bar{\sigma}=\begin{pmatrix}\sigma_{a*a*} & 0 & \sigma_{a*c}\\
0 & \sigma_{bb} & 0\\
\sigma_{a*c} & 0 & \sigma_{cc}
\end{pmatrix}\label{a}
\end{equation}
where the off-diagonal elements both have the same value $\sigma_{a*c}$.
For isotropic materials, the conductivity tensor is a unity matrix
times the scalar conductivity value. Different values of the diagonal
elements indicate different conductivity values along the axes of
the coordinate system. Off-diagonal elements indicate a rotation between
the axes of the coordinate system and the directions of minimum and
maximum conductivity.

\subsection{Calculation of the 2-dimensional in-plane conductivity tensor}

\begin{figure}
\includegraphics[scale=0.6]{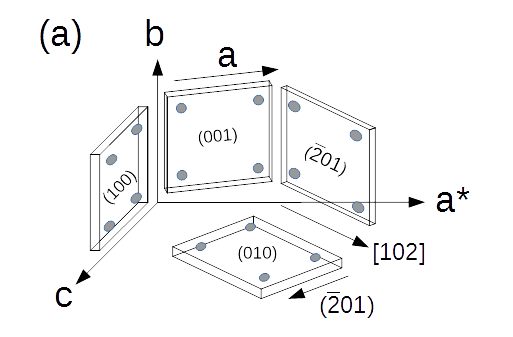} \includegraphics[scale=0.6]{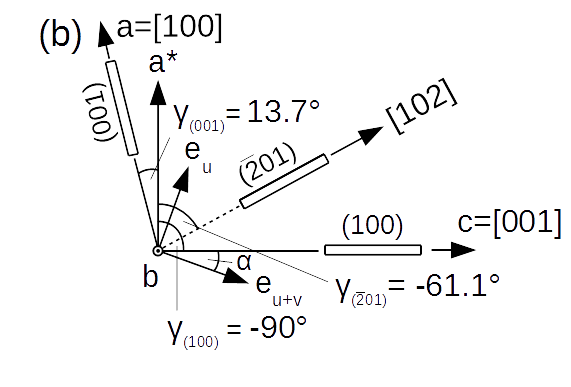}
\caption{\label{sview}(a) Overview of the orientation of the samples. Three
samples have one edge along the b-direction and the other edge is
within the $ac$ plane. (b) These three samples and the angles between
them are shown in a projection onto the $ac$-plane. This sketch includes
the room-temperature directions of minimum and maximum conductivity
in the $ac$ plane.}
\end{figure}

Measurements using the van der Pauw configuration probe the conduction
in the plane of the two-dimensional (2D) sample. Thus, a two-dimensional
conductivity tensor is required for the analysis and the modeling
of the experiment. The non-trivial relation between 2D and 3D tensor
for monoclinic materials will be described next for the major available
substrate orientations of $\beta$-Ga$_{2}$O$_{3}$ including those
used in this work. An overview of these orientations is presented
in Fig.~\ref{sview}.

Using the conductivity tensor $\bar{\sigma}$ given in Eq.~\ref{a},
the relation $\vec{j}=\bar{\sigma}\vec{E}$ can be written as:

\begin{equation}
\begin{pmatrix}j_{a*}\\
j_{b}\\
j_{c}
\end{pmatrix}=\bar{\sigma}\begin{pmatrix}E_{a*}\\
E_{b}\\
E_{c}
\end{pmatrix}=\begin{pmatrix}\sigma_{a*a*}E_{a*}+\sigma_{a*c}E_{c}\\
\sigma_{bb}E_{b}\\
\sigma_{a*c}E_{a*}+\sigma_{cc}E_{c}
\end{pmatrix}\label{jsejs0}
\end{equation}

in the $a${*}$bc$ Cartesian coordinate system. This equation will
be used next for the determination of the 2D conductivity tensor in
terms of 3D conductivity tensor elements. The difference compared
to isotropic materials can be seen in Eq.~\ref{jsejs0}: the direction
of $\vec{E}$ can be different from the direction of the current $\vec{j}$,
i. e. an out of plane electric field can exist for samples with in-plane
current and the orientation of the sample compared to the reference
system has to be considered.

\subsubsection{(100) orientation}

For our (100)-surface samples (=$a${*}) with edges along {[}010{]}=$b$
and {[}001{]}=$c$, defining our Cartesian a{*}bc reference system,
the condition of zero current perpendicular to the surface ($j_{a*}=j_{\perp}=0$),
can be applied to Eq.~\ref{jsejs0}: 
\begin{equation}
0=\sigma_{a*a*}E_{a*}+\sigma_{a*c}E_{c}\label{e1e3001}
\end{equation}
which is equivalent to fixing the electric field perpendicular to
the surface to 
\begin{equation}
E_{a*}=-\frac{\sigma_{a*c}E_{c}}{\sigma_{a*a*}}
\end{equation}
Using Eq.~\ref{jsejs0} and Eq.~\ref{e1e3001}: 
\begin{eqnarray}
\begin{pmatrix}j_{a*}\\
j_{b}\\
j_{c}
\end{pmatrix} & = & \begin{pmatrix}\sigma_{a*a*}(-\frac{\sigma_{a*c}E_{c}}{\sigma_{a*a*}})+\sigma_{a*c}E_{c}\\
\sigma_{bb}E_{b}\\
\sigma_{a*c}(-\frac{\sigma_{a*c}E_{c}}{\sigma_{a*a*}})+\sigma_{cc}E_{c}
\end{pmatrix}\nonumber \\
 & = & \begin{pmatrix}0\\
\sigma_{bb}E_{b}\\
(\sigma_{cc}-\frac{\sigma_{a*c}^{2}}{\sigma_{a*a*}})E_{c}
\end{pmatrix}
\end{eqnarray}
which can be rewritten to 
\begin{equation}
\begin{pmatrix}j_{a*}\\
j_{b}\\
j_{c}
\end{pmatrix}=\begin{pmatrix}0 & 0 & 0\\
0 & \sigma_{bb} & 0\\
0 & 0 & \sigma_{cc}-\frac{\sigma_{a*c}^{2}}{\sigma_{a*a*}}
\end{pmatrix}\begin{pmatrix}E_{a*}\\
E_{b}\\
E_{c}
\end{pmatrix}\mbox{\,.}\label{3d2d}
\end{equation}
The 2D tensor in the coordinate system of the sample edges, the axes
$b$ and $c$, is now given by the $bc$-components of the conductivity
tensor in Eq. \ref{3d2d}: 
\begin{equation}
\bar{\sigma}^{bc}=\begin{pmatrix}\sigma_{bb} & 0\\
0 & \sigma_{cc}-\frac{\sigma_{a*c}^{2}}{\sigma_{a*a*}}
\end{pmatrix}\label{(100)s2d}
\end{equation}
The experimentally determined value of the conductivity anisotropy
of one sample can now be given as the ratio of the diagonal elements
in Eq.~\ref{(100)s2d}: 
\begin{equation}
A_{\sigma}^{(100)}=\frac{\sigma_{cc}-\frac{\sigma_{a*c}^{2}}{\sigma_{a*a*}}}{\sigma_{bb}}\label{eq:A100}
\end{equation}

\subsubsection{(001) orientation}

The same derivation as for the (100)-surface samples can be done for
(001)-surface samples(=$c${*}) (edges along {[}100{]}=$a$ and {[}010{]}=$b$),
but in the $abc${*} Cartesian coordinate system. The angle between
the $a${*}$bc$ and $abc${*} Cartesian coordinate systems is 13.7$^{\circ}$
around the $b$-axis (=angle between $a$ and $a${*} as well as $c${*}
and $c$). 
\begin{equation}
\bar{\sigma}^{ab}=\begin{pmatrix}\sigma_{aa}-\frac{\sigma_{ac*}^{2}}{\sigma_{c*c*}} & 0\\
0 & \sigma_{bb}
\end{pmatrix}\label{(001)s2d}
\end{equation}
whose components consist of conductivity components of the 3D conductivity
tensor written in the $abc${*} system. Hence, the experimentally
determined value of the conductivity anisotropy is the ratio of the
diagonal elements in Eq.~\ref{(001)s2d}: 
\begin{eqnarray}
A_{\sigma}^{(001)} & = & \frac{\sigma_{aa}-\frac{\sigma_{ac*}^{2}}{\sigma_{c*c*}}}{\sigma_{bb}}\mbox{\,.}\label{eq:A001}
\end{eqnarray}

To relate the $abc${*}-system tensor components to the $a${*}$bc$-system
tensor components of Eq.~\ref{a}, a rotation has to be applied to
the tensor. 
\begin{equation}
\bar{\sigma}_{rot}=R_{\alpha}\bar{\sigma}R_{\alpha}^{\text{-1}}\label{eq:srsr}
\end{equation}
where the rotated conductivity tensor $\sigma_{rot}$ follows a unitary
transformation using the rotation matrix for a rotation around the
$b$-axis 
\begin{equation}
R_{\alpha}=\begin{pmatrix}\text{cos}(\alpha) & 0 & \text{sin}(\alpha)\\
0 & 1 & 0\\
-\text{sin}(\alpha) & 0 & \text{cos}(\alpha)
\end{pmatrix}
\end{equation}
The transformation described by Eq.~\ref{eq:srsr} can be derived
by writing the relation $\vec{j}=\bar{\sigma}\vec{E}$ before and
after applying a rotation: 
\begin{equation}
\vec{j}_{rot}=R_{\alpha}\vec{j}=R_{\alpha}\bar{\sigma}\vec{E}=R_{\alpha}\bar{\sigma}R_{\alpha}^{\text{-1}}R_{\alpha}\vec{E}=\bar{\sigma}_{rot}\vec{E}_{rot}
\end{equation}
To rotate the coordinate system by a certain angle, both vectors and
matrices have to be rotated by the negative value of that angle. To
transform a tensor from the $abc${*}-system to the $a${*}$bc$-system,
a rotation around the $b$-axis of $\gamma_{(001)}=-13.7^{\circ}$
has to be applied to the tensor. Using this rotation, the $abc${*}
tensor elements can be expressed in terms of the $a${*}$bc$ tensor
elements: 
\begin{eqnarray}
\sigma_{aa} & = & \sigma_{a*a*}\text{cos}^{2}\gamma_{(001)}-\text{sin}(2\gamma_{(001)})\sigma_{a*c}\nonumber \\
 &  & +\sigma_{cc}\text{sin}^{2}\gamma_{(001)}\label{eq:001rot}\\
\sigma_{ac*} & = & \text{sin}\gamma_{(001)}\text{cos}\gamma_{(001)}\sigma_{a*a*}+\text{cos}(2\gamma_{(001)})\sigma_{a*c}\nonumber \\
 &  & -\text{sin}\gamma_{(001)}\text{cos}\gamma_{(001)}\sigma_{cc}\\
\sigma_{c*c*} & = & \text{sin}^{2}\gamma_{(001)}\sigma_{a*a*}+\text{sin}(2\gamma_{(001)})\sigma_{a*c}\nonumber \\
 &  & +\text{cos}^{2}\gamma_{(001)}\sigma_{cc}\label{eq:001rot3}
\end{eqnarray}
where $\sigma_{bb}$ remains unchanged.

\subsubsection{($\bar{2}\,0\,1$) orientation}

For ($\bar{2}\,0\,1$)-sample, the 2D conductivity tensor in the coordinate
system can be derived like for the (1~0~0)- and the (0~0~1)-surface
samples, but in the coordinate system of the sample edges {[}0~1~0{]}
and {[}1~0~2{]}. In that system, the two-dimensional conductivity
tensor can be given as: 
\begin{equation}
\bar{\sigma}^{[102],b}=\begin{pmatrix}\sigma_{[102][102]}-\frac{\sigma_{(\bar{2}01)[102]}^{2}}{\sigma_{(\bar{2}01)(\bar{2}01)}} & 0\\
0 & \sigma_{bb}
\end{pmatrix}\label{(201)s2d}
\end{equation}
Hence, the experimentally determined value of the conductivity anisotropy
is the ratio of the diagonal elements in Eq.~\ref{(201)s2d}:\\

\begin{eqnarray}
A_{\sigma}^{(\bar{2}01)} & = & \frac{\sigma_{[102][102]}-\frac{\sigma_{(\bar{2}01)[102]}^{2}}{\sigma_{(\bar{2}01)(\bar{2}01)}}}{\sigma_{bb}}\mbox{\,.}\label{eq:A201}
\end{eqnarray}
Using a rotation of $\gamma_{(\bar{2}01)}=61.1^{\circ}$ around the
$b$-axis, the $[1\,0\,2]b(\bar{2\,}0\,1)$ tensor elements can be
expressed in terms of the $a${*}$bc$ tensor elements in the same
form as for the (0~0~1)-surface samples:

\begin{eqnarray}
\sigma_{[102][102]} & = & \text{cos}^{2}\gamma_{(\bar{2}01)}\sigma_{a*a*}-\text{sin}(2\gamma_{(\bar{2}01)})\sigma_{a*c}\nonumber \\
 &  & +\text{sin}^{2}\gamma_{(\bar{2}01)}\sigma_{cc}\label{eq:201rot}\\
\sigma_{[102](\bar{2}01)} & = & \text{sin}\gamma_{(\bar{2}01)}\text{cos}\gamma_{(\bar{2}01)}\sigma_{a*a*}+\text{cos}(2\gamma_{(\bar{2}01)})\sigma_{a*c}\nonumber \\
 &  & -\text{sin}\gamma_{(\bar{2}01)}\text{cos}\gamma_{(\bar{2}01)}\sigma_{cc}\\
\sigma_{(\bar{2}01)(\bar{2}01)} & = & \text{sin}^{2}\gamma_{(\bar{2}01)}\sigma_{a*a*}+\text{sin}(2\gamma_{(\bar{2}01)})\sigma_{a*c}\nonumber \\
 &  & +\text{cos}^{2}\gamma_{(\bar{2}01)}\sigma_{cc}\label{eq:201rot3}
\end{eqnarray}

where $\sigma_{bb}$ again remains unchanged.

\subsubsection{(010) orientation}

For completeness we are giving the transformation for (0~1~0)-surface
samples with edges along {[}1~0~2{]} and $\perp[1\,0\,2]$ as well.
The three-dimensional conductivity tensor from Eq.~\ref{a} can be
directly rewritten into a two-dimensional Tensor in the a{*}c-system
by removing the second row and column: 
\begin{equation}
\sigma^{a*c}=\begin{pmatrix}\sigma_{a*a*} & \sigma_{a*c}\\
\sigma_{a*c} & \sigma_{cc}
\end{pmatrix}
\end{equation}
As the sample edges are along {[}$1\,0\,2${]} and $\perp[1\,0\,2]$,
a rotation of about $\gamma_{(010)}=61.1^{\circ}$ (=$\gamma_{(\bar{2}01)}$)
around the $b$-axis has to be applied to get the conductivity tensor
in the coordinate system of the sample edges: 
\begin{equation}
\sigma^{[102],\perp[102]}=\begin{pmatrix}\sigma_{[102][102]} & \sigma_{[102],\perp[102]}\\
\sigma_{[102,\perp[102]} & \sigma_{\perp[102],\perp[102]}
\end{pmatrix}
\end{equation}
where: 
\begin{eqnarray}
\sigma_{[102][102]} & = & \text{cos}^{2}\gamma_{(010)}\sigma_{a*a*}-\text{sin}(2\gamma_{(010)})\sigma_{a*c}\\
 &  & +\text{sin}^{2}\gamma_{(010)}\sigma_{cc}\nonumber \\
\sigma_{[102],\perp[102]} & = & \text{sin}\gamma_{(010)}\text{cos}\gamma_{(010)}\sigma_{a*a*}+\text{cos}(2\gamma_{(010)})\sigma_{a*c}\nonumber \\
 &  & -\text{sin}\gamma_{(010)}\text{cos}\gamma_{(010)}\sigma_{cc}\\
\sigma_{\perp[102],\perp[102]} & = & \text{sin}^{2}\gamma_{(010)}()\sigma_{a*a*}+\text{sin}(2\gamma_{(010)})\sigma_{a*c}\nonumber \\
 &  & +\text{cos}^{2}\gamma_{(010)}\sigma_{cc}\mbox{\,.}
\end{eqnarray}

\subsection{Reconstructing the 3-dimensional conductivity tensor}

The 3D conductivity tensor has four independent elements (compare
Eq.~\ref{a}). Measurements of the 2D in-plane conductivity anisotropy
of three differently oriented samples are thus sufficient to characterize
the 3D conductivity anisotropy, as the tensor components can be normalized
to one of the components. The (100), the (001) and the ($\bar{2}01$)
surface samples all share one edge along the $b$-direction and the
other one in the $ac$ plane (visualized in Fig.~\ref{sview}(a)).
Thus, the measured anisotropies (Eqs.~\ref{eq:A100},~\ref{eq:A001},~\ref{eq:A201})
are ratios of $ac$-plane conductivities and $\sigma_{bb}$.

The 3D conductivity tensor Eq.~\ref{a} can be diagonalized by rotating
the $a${*}$bc$ coordinates system by an angle $\alpha$ around the
$b$-axis into an arbitrary coordinate system, whose axes in the $ac$-plane
are given by the directions $e_{u}$ and $e_{u+v}$ of minimum and
maximum conductivity $u$ and ($u$+$v$), respectively, in the $ac$-plane
(as illustrated in Fig.~\ref{sview}(b)): 
\begin{equation}
\sigma^{min,max}=\begin{pmatrix}u & 0 & 0\\
0 & \sigma_{bb} & 0\\
0 & 0 & u+v
\end{pmatrix}\mbox{\,.}\label{uvtensor}
\end{equation}

Conversely, $\sigma^{min,max}$ can be expressed in the $a${*}$bc$
system by a rotation of the coordinate system of $-\alpha$ around
the $b$-axis: 
\begin{eqnarray}
\sigma^{a*bc} & = & \begin{pmatrix}\sigma_{a*a*} & 0 & \sigma_{a*c}\\
0 & \sigma_{bb} & 0\\
\sigma_{a*bc} & 0 & \sigma_{cc}
\end{pmatrix}\nonumber \\
 & = & \begin{pmatrix}u+v\text{sin}^{2}\alpha & 0 & v\text{sin}\alpha\text{cos}\alpha\\
0 & \sigma_{bb} & 0\\
v\text{sin}\alpha\text{cos}\alpha & 0 & u+v\text{cos}^{2}\alpha
\end{pmatrix}\mbox{\,.}\label{eq:sigmacstaruvalpha}
\end{eqnarray}

Using the relation between the $e_{u}be_{u+v}$ and the $a${*}$bc$
coordinate system conductivity values in Eq.~\ref{eq:sigmacstaruvalpha},
Eq.~\ref{eq:A100} can be rewritten as 
\begin{equation}
A_{\sigma}^{(100)}=\frac{(u+v\text{cos}^{2}\alpha)-\frac{(v\text{sin}\alpha\text{cos}\alpha)^{2}}{(u+v\text{sin}^{2}\alpha)}}{\sigma_{bb}}\label{eq:A100uv}
\end{equation}
which is equivalent to 
\begin{equation}
A_{\sigma}=\frac{u/\sigma_{bb}+v/\sigma_{bb}}{1+\frac{v}{u}\sin^{2}(\alpha)}
\end{equation}

Using $\gamma$, the angle describing the orientation of the sample
edges in the $a${*}$bc$ coordinate system (see Fig.~\ref{sview}(b)),
that equation can be rewritten to 
\begin{equation}
\frac{u/\sigma_{bb}+v/\sigma_{bb}}{1+\frac{v}{u}\cos^{2}(\alpha-\gamma)}=A_{\sigma}\label{eq:Asigmauvbb}
\end{equation}
with $\gamma_{(100)}$=-90°. The same can be done for the (001) and
the ($\bar{2}01$)-surface samples, taking Eqs.~\ref{eq:A001}~and~\ref{eq:A201}
to arrive at the same equation. The rotation between the system of
the sample edges and the $a${*}$bc$-system (Eqs.~\ref{eq:001rot}~to~\ref{eq:001rot3}
and~\ref{eq:201rot}~to~\ref{eq:201rot3}) is expressed in $\gamma$
in each case. For our oriented samples, $\gamma_{(001)}=13.7\text{°}$
and $\gamma_{(\bar{2}01)}=-61.1\text{°}$.

For these three sample orientations $\alpha$, $u/\sigma_{bb}$, and
$v/\sigma_{bb}$ are the same, and can thus be determined by solving
the resulting system of the three coupled Eqs.~\ref{eq:Asigmauvbb}
numerically using the experimentally obtained $A_{\sigma}$ and known
$\gamma$. The conductivity tensor in the a{*}c system can then be
readily calculated from these results by equating the tensor components
in Eq.~\ref{eq:sigmacstaruvalpha}.

\section{results and discussion}

\subsection{Room temperature results}

\begin{table}[h]
\caption{Values of the measured 2D conductivity anisotropy $A$ at room temperature
for different samples. The values were derived from comparing the
measured van der Pauw anisotropy $A_{vdP}$ to finite element simulations
(see Tab.~\ref{tab:AvsAvdP}). As the off-diagonal element is much
smaller than the diagonal elements, the given values are an approximation
for the ratio of the conductivity along one side of the sample ($c$-axis,
$a$-axis, $[1~0~2]$) and the other ($b$-axis).}
\begin{tabular}{llr}
\hline 
Sample  & 2D anisotropy  & 2D anisotropy\tabularnewline
name  & (3D Tensor elements)  & value\tabularnewline
\hline 
$G100a$  & $\frac{\sigma_{cc}-\frac{\sigma_{a*c}^{2}}{\sigma_{a*a*}}}{\sigma_{bb}}$  & $0.960\pm0.01$\tabularnewline
$G100b$  & $\frac{\sigma_{cc}-\frac{\sigma_{a*c}^{2}}{\sigma_{a*a*}}}{\sigma_{bb}}$  & $1.000\pm0.02$\tabularnewline
$G100c$ (extended defects)  & $\frac{\sigma_{cc}-\frac{\sigma_{a*c}^{2}}{\sigma_{a*a*}}}{\sigma_{bb}}$  & $0.768\pm0.01$\tabularnewline
$G001a$  & $\frac{\sigma_{aa}-\frac{\sigma_{ac*}^{2}}{\sigma_{c*c*}}}{\sigma_{bb}}$  & $1.000\pm0.01$\tabularnewline
$G-201a$  & $\frac{\sigma_{[102]}-\frac{\sigma_{[102](\bar{2}01)}^{2}}{\sigma_{(\bar{2}01)}}}{\sigma_{bb}}$  & $0.995\pm0.01$\tabularnewline
\hline 
\end{tabular}\label{tabres} %
\end{table}
The measured value of the 2D conductivity anisotropy at room temperature,
together with their relation to the three-dimensional conductivity
tensor $\bar{\sigma}$ by Eqs.~\ref{(100)s2d}, \ref{(001)s2d},
and \ref{(201)s2d} are summarized in Tab.~\ref{tabres}. For all
samples, except G100c containing extended defects, a fairly isotropic
in-plane conductivity was observed at room temperature with deviation
of the conductivity anisotropy from unity of below $5\%$ and even
below $1\%$ for three of these samples. Theory suggests a dependence
of transport anisotropy on electron concentration or prevalent scattering
mechanism (see Refs.\,\cite{Kang,Ghosh2017} and Tab.\,\ref{tablit}).
We are thus using the data of samples G100a and G001a, which have
very similar electron concentrations, and that of G-201a to obtain
the conductivity tensor Eq.~\ref{a} in the $a${*}$bc$ reference
system by solving the system of coupled Eqs.~\ref{eq:Asigmauvbb}
. The resulting tensor (normalized to the conductivity in $b$-direction)
at room temperature and an electron concentration of $n=(3.5\pm1)\times10^{17}$\,cm$^{-3}$
is: %
\begin{equation}
\bar{\sigma}/\sigma_{bb}(T=300\,K)=\begin{pmatrix}1.02\pm0.02 & 0 & 0.03\pm0.03\\
0 & 1 & 0\\
0.03\pm0.03 & 0 & 0.96\pm0.02
\end{pmatrix}.\label{s-tensor}
\end{equation}
Our results indicate a small anisotropy of the conductivity with $\sigma_{a*a*}$
and $\sigma_{cc}$ being few \% higher and lower than $\sigma_{bb}$,
respectively, and the non-diagonal elements $\sigma_{a*c}$ amounting
to no more than a few~\% of $\sigma_{bb}$.

To analyze the anisotropies of the scattering time, the conductivity
tensor can be compared to an effective mass tensor. The tensor taken
from \cite{Furthmuller}, which Furthmuller et al. calculated by first
principles, can be rotated to the $a${*}$bc$ reference system and
normalized to $m${*} along the $b$-axis: 
\begin{equation}
\bar{m^{*}}/m_{bb}^{*}=\begin{pmatrix}1.009 & 0 & -0.001\\
0 & 1 & 0\\
-0.001 & 0 & 0.973
\end{pmatrix}\label{m-tensor}
\end{equation}
As the conductivity and the effective mass are close to isotropic,
the overall scattering time at room temperature is quite isotropic
as well. Using the relation $\bar{\sigma}=e\frac{\bar{\tau}}{\bar{m^{*}}}$,
the results from Eq.~\ref{s-tensor}, and the effective mass tensor
from \cite{Furthmuller}, the $\tau$-tensor normalized to $\tau_{bb}$,
can be calculated to be: 
\begin{equation}
\bar{\tau/\tau_{bb}}=\begin{pmatrix}1.03\pm0.02 & 0 & 0.03\pm0.03\\
0 & 1 & 0\\
0.03\pm0.03 & 0 & 0.93\pm0.02
\end{pmatrix}
\end{equation}
at room temperature with $\tau_{a*a*}$ being 10\% higher than $\tau_{cc}$.

\subsection{Temperature dependence and scattering mechanisms}

\begin{figure}
\includegraphics[width=8cm]{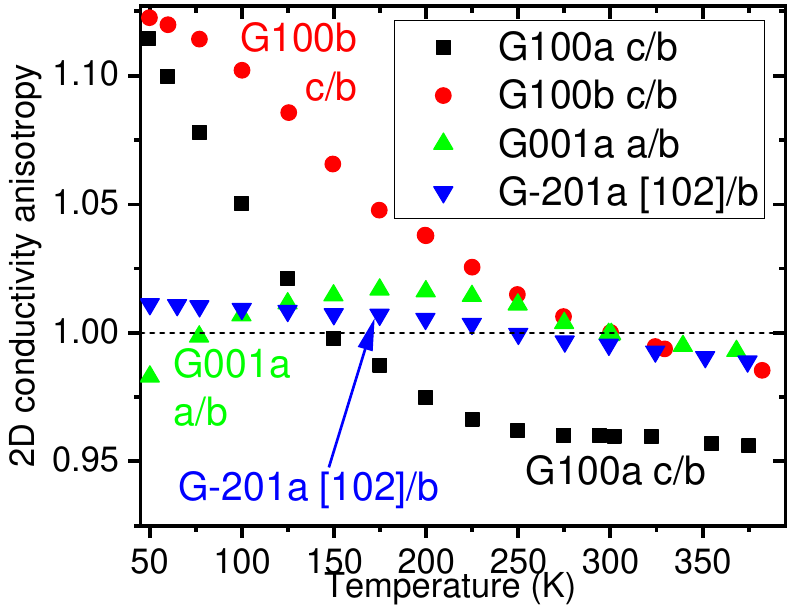}\caption{Temperature-dependent 2D conductivity anisotropy $A$ for samples
with different surface orientations. The in-plane directions of the
sample edges that result in the conductivity anisotropy, e.g. ``c/b''
resulting in $A=\sigma_{\parallel c}/\sigma_{\parallel b}$, are given
with the sample name. The dashed line denotes completely isotropic
in-plane conductivity.}
\label{aniso2} 
\end{figure}
The temperature dependence between 50\,K and 380\,K of measured
2D-conductivity anisotropies $A$ in the coordinate system of the
sample edges are shown in Fig.~\ref{aniso2}. Samples G001a and G-201a
remain completely isotropic within 2\% and 1\%, respectively, for
the entire temperature range. For the temperature decreasing from
$\approx370$\,K to $50$\,K both, G100a and G100b, clearly show
an increasing anisotropy ($A=\sigma_{\parallel c}/\sigma_{\parallel b}$)
from $0.96$ and $0.99$ to $1.11$ and $1.12$, respectively. A deviation
of $A$ from unity of less than than 5\% can be seen for temperatures
between 175~K and 380~K for all samples. These results suggest a
rather isotropic conductivity at application-relevant (high) temperatures.

\begin{figure}
\includegraphics[width=4.2cm]{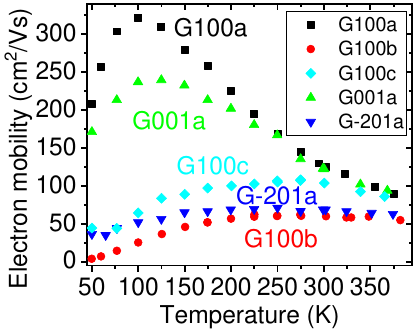}\includegraphics[width=4.2cm]{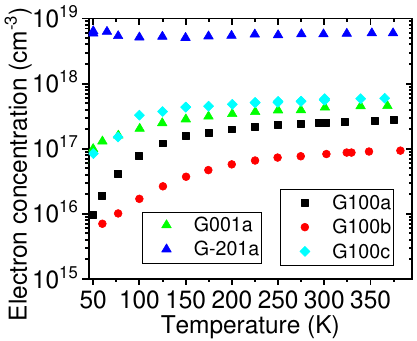}
\caption{Average Hall mobility (left) and electron concentration (right) as
a function of temperature for various sample orientations. Measured
values were corrected for geometry effects of non-edge contact placement
and finite contact size (see Tab.\,\ref{tabfem2} for correction
factors).}
\label{moba-ccd} 
\end{figure}

Next, we will address the impact of scattering mechanism on transport
anistropy. The electron mobility in $\beta$-Ga$_{2}$O$_{3}$ free
from extended defects is mainly limited by polar longitudinal optical
phonon scattering (PLOPS) and ionized impurity scattering (IIS) at
fixed ionized point charges\cite{Ma,Kang}. With increasing temperature
the electron mobility limited by PLOPS or IIS decreases or increases,
respectively, and higher ionized point charge concentrations (including
dopants, compensating dopants, or point defects) decrease the IIS-limited
mobility\cite{Ma}. To identify the dominant scattering mechanisms
(PLOPS or IIS) for all samples we compare our measured temperature
dependent mobilities and electron concentrations shown in Fig.~\ref{moba-ccd}
to modeled- and reference data from Ref.~\cite{Ma}: The high mobility
at $T=300$~K which strongly increases with decreasing temperature
in samples G100a and G001a (Fig.~\ref{moba-ccd}, left) indicates
dominant PLOPS at $T\gtrsim200$~K. Consistent with this assignment,
the intermediate electron concentrations of these samples (Fig.~\ref{moba-ccd},
right) corresponds to a donor concentration that is low enough (and
not significantly compensated) to be clearly in the regime of dominant
PLOPS (cf. Ref.~\cite{Ma}). Due to the small off-diagonal element
$\sigma_{a*c}/\sigma_{bb}$, the experimentally obtained conductivity
anisotropies of G100a and G001a shown in Fig.~\ref{aniso2}, are
good approximations of $\sigma_{cc}/\sigma_{bb}\approx0.96$ and $\sigma_{aa}/\sigma_{bb}\approx1.00$
for PLOPS. This fairly isotropic PLOPS is in fair agreement with the
theoretically predicted $\sigma_{c*c*}/\sigma_{bb}\approx1.02$ of
Ref.~\cite{Kang} but significantly contrasts the predicted stronger
anisotropy of $\sigma_{c*c*}/\sigma_{bb}\approx0.64$ of Ref.\,\cite{Ghosh2017}
and $\sigma_{aa}/\sigma_{bb}\approx0.84$ of Ref.~\cite{Kang} for
comparable electron concentrations. (We note that $a$ and $a${*}
as well as $c${*} and $c$ are different from each other. The rotation
from the $abc${*}-system (used in the theory Refs.~\cite{Kang,Ghosh2017})
to the the $a${*}$bc$-system (used by us) by 13.7°, however, is
small enough to grant quantitative comparison of the anisotropies
along these directions.)
In marked contrast, the mobilities of G-201a and G100b are significantly
lower and show only a weak temperature dependence at $T\gtrsim200$~K,
indicating strong IIS. The high, degenerate electron concentration
shown in Fig.~\ref{moba-ccd},\,right for G-201a, implying a high
donor concentration, corroborates strong IIS, whereas strong IIS in
G100b with rather low electron concentration can only be explained
by compensation. Our experimental results further limit the uncertainty,
in particular with regard to the directions, from a maximum 40\% and
10\% deviation from the isotropic case experimentally determined by
optical Hall effect measurements of samples with high donor concentration\cite{Knight}
and estimated theoretically for IIS\cite{Kang}, respectively: Comparing
G100b (strong IIS) to G100a (weaker IIS) we find the anisotropy $A=\sigma_{\parallel c}/\sigma_{\parallel b}\approx\sigma_{cc}/\sigma_{bb}$
to be systematically higher for G100b (see in Fig.\,\ref{aniso2}).
This behavior is in qualitative agreement with an increase of $A$
for G100a with decreasing temperature, i.e. with decreasing PLOPS
and increasing IIS. At $T\lesssim100$~K decreasing mobility with
decreasing temperature indictes dominant IIS in all samples. Under
these conditions the anisotropy of G100a,b reaches $\sigma_{cc}/\sigma_{bb}\approx1.12$
whereas $\sigma_{[102]}/\sigma_{bb}$ of G-201a and $\sigma_{aa}/\sigma_{bb}$
of G001a remain within 1\% and 2\% around unity. These results indicate
fairly isotropic IIS. The systematically higher conductivity in the
$c$-direction compared to the almost equal conductivities in the
$a-$ and $b-$direction is likely related to higher dielectric constant
for the $c$-direction compared to the almost equal dielectric constants
for directions perpendicular to $c$.\cite{Ghosh2016,Fiedler2019}%
{} 

\begin{figure}
\includegraphics[width=8cm]{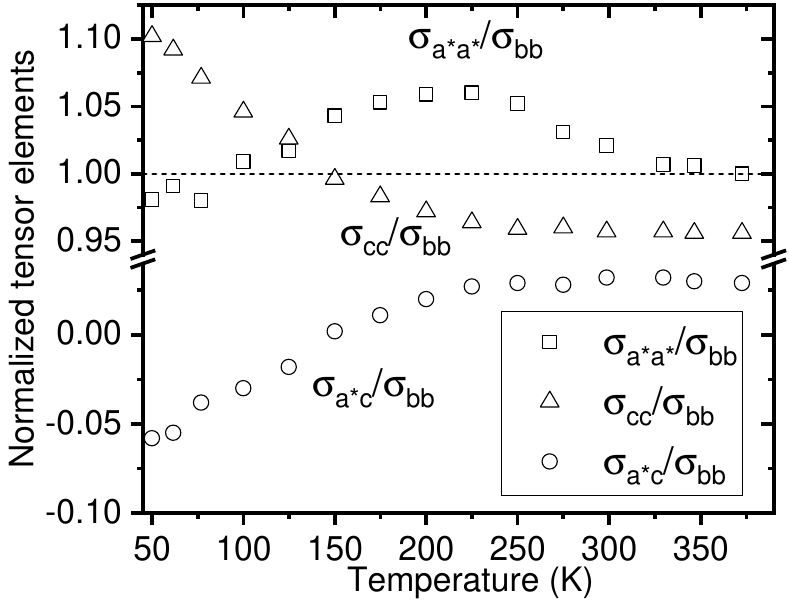} \caption{Visualization of the temperature-dependent 3D conductivity tensor
elements of $\text{\ensuremath{\beta}-Ga}_{\text{2}}\text{O}_{\text{3}}$
normalized to the conductivity in $b$-direction, $\sigma_{bb}$.}
\label{telement} 
\end{figure}
Neglecting potentially different anisotropies associated to different
scattering mechanisms (in samples G-201a compared to G100a and G001a)
we can determine the temperature-dependent conductivity tensor Eq.~\ref{a}
normalized to the conductivity in $b$-direction, $\bar{\sigma}/\sigma_{bb}$,
by numerically solving Eq.~\ref{eq:Asigmauvbb} with the data from
our experimentally temperature-dependent 2D conductivity anisotropies
of G100a, G001a, and G-201a. The resulting temperature dependent tensor
components are shown in Fig.~\ref{telement}.

\subsection{Anisotropy by extended defects}

\begin{figure}
\includegraphics[width=8cm]{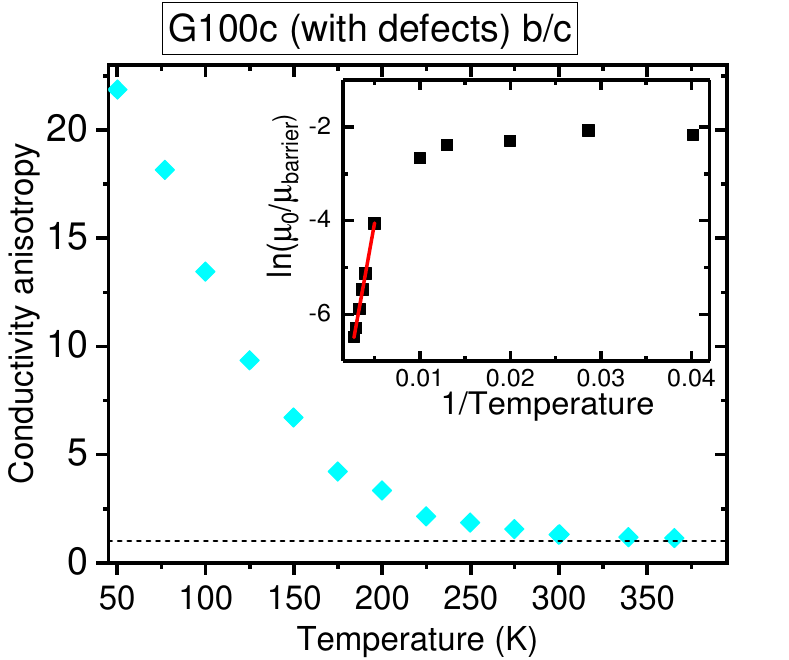} \caption{\label{iso18log}Conductivity anisotropy of sample G100c containing
visible defects (low-angle grain boundaries along the $b$-direction).
The dashed line denotes completely isotropic in-plane conductivity.
The inset shows an Arrhenius plot of the barrier-related mobility,
the fit between 200~K and 365~K results in an activation energy
of $38\pm2$~meV.}
\end{figure}

Large conductivity anisotropies with significantly higher conductivity
in the $b$ than in the $c$-direction have been identified in sample
G100c, which contains low-angle grain boundaries oriented approximately
along the $b$-direction (see and Fig.~\ref{go3}). In this sample,
the room temperature conductivity anisotropy $\sigma_{cc}/\sigma_{bb}\approx0.77$
(shown in Tab.~\ref{tabres}) is much stronger than that of the comparable
high-quality samples G100a and G100b. A qualitatively similar behavior
with $\sigma_{cc}/\sigma_{bb}\approx0.06$ and $\sigma_{cc}/\sigma_{bb}\approx0.5$
at room temperature has been observed in $\text{\ensuremath{\beta}-Ga}_{\text{2}}\text{O}_{\text{3}}$
(100) bulk samples of unknown structural quality\cite{Ueda} and MOVPE
layers containing a high density of incoherent twin boundaries oriented
along the $b$-direction.\cite{Fiedler} Planar defects, such as twin-
or grain boundaries are typically associated with energy barriers
that impede electron transport across, explaining the relatively lower
conductivity along the $c$-direction for G100c\cite{Seto,OrtonPowell}.
For convenience we plot the temperature dependence of the inverse
anisotropy, $\sigma_{bb}/\sigma_{cc}$, for sample G100c in Fig.~\ref{iso18log}.
A strong increase of anisotropy with decreasing temperature from $\sigma_{bb}/\sigma_{cc}=1.14$
at $T=360\,K$ to $\sigma_{bb}/\sigma_{cc}>20$ at $T=50\,K$ is observed.
It follows an activated behavior and the activation energy for the
conductivity in the $c$ direction (across the barriers) for $T>100$~K
was calculated according to $\sigma_{[001]}=\sigma_{0}e^{-\frac{E_{a}}{kT}}$
to be $E_{A}=(38\pm2)$~meV. (We note that the increasing anisotropy
$\sigma_{cc}/\sigma_{bb}$ with decreasing temperature in samples
G100a and G100b cannot be related to such extended defects as it is
inverse to the increasing $\sigma_{bb}/\sigma_{cc}$ of G100c.)

\begin{figure}
\includegraphics[width=8cm]{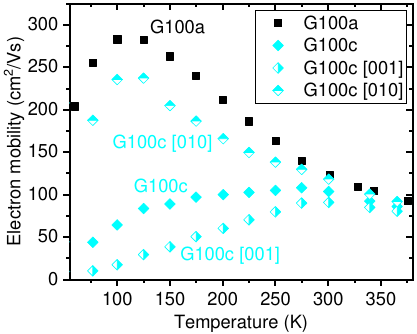} \caption{\label{mob18}Temperature-dependent Hall mobility for sample G100c
containing low-angle grain boundaries along the $b$-direction. Cyan
rhombs: average mobility and mobility along and perpendicular to the
defects. Black squares: mobility of the sample G100a without these
defects for comparison.}
\end{figure}
Figure~\ref{mob18} shows the mobility in the direction parallel
and perpendicular to the low-angle grain boundary defects in sample
G100c. Parallel to the defects, the mobility is only slightly lower
than that of a comparable sample without these defects, G100a, whereas
it is drastically reduced perpendicular to the defects, especially
at low temperatures. Above room temperature, however, the mobilities
in both direction coincide and decrease with increasing temperature
as expected for phonon-limited transport, indicating a negligible
effect of the grain boundaries. Assuming the mobility $\mu$ within
the grains to be isotropic the total mobility perpendicular to the
defects ({[}001{]}-direction), $\mu_{\perp}$, is reduced due to the
activated, barrier-related mobility $\mu_{barrier}$ : 
\begin{equation}
\frac{1}{\mu_{\perp}}=\frac{1}{\mu}+\frac{1}{\mu_{barrier}}
\end{equation}
As $A_{\sigma}\approx\frac{\sigma_{bb}}{\sigma_{cc}}\approx\mu/\mu_{\perp}$:
\begin{equation}
\mu_{barrier}=\frac{\mu}{A_{\sigma}-1}=\mu_{0}e^{\frac{E_{a}}{kT}}
\end{equation}
using this equation, the activation energy for the transport across
the barriers is $E_{A}=(93\pm2)$~meV for temperatures between $T=200$~K
and $T=365$~K (see inset of Fig.~\ref{iso18log}). Interestingly,
this barrier is orders of magnitude higher than that found at low-angle
grain boundaries of bulk SnO$_{2}$ \cite{BierwagenGalazka}, which
might also explain the difficulties in realizing conductive, non-single
crystalline $\beta$-Ga$_{2}$O$_{3}$ films. 

\section{Summary and Conclusion}

Using van der Pauw measurements with well-defined contact geometry
and careful FEM modeling of the structures, the anisotropy of the
electrical conductivity in high-quality bulk $\text{\ensuremath{\beta}-Ga}_{\text{2}}\text{O}_{\text{3}}$
wafers with different surface orientation was precisely determined.

Most importantly, fairly isotropic behavior with conductivity ratios
of different directions between $0.96\pm0.01$ and $1.06\pm0.02$
at temperatures between 150~K and 375~K were found at electron concentrations
on the order of $10^{17}$\,cm$^{-3}$. Based on the extracted anisotropies,
the ratio of the non-zero elements of the conductivity tensor were
determined. Comparison to the effective mass tensor indicates that
the transport anisotropy is strongly influenced by the scattering
times of anisotropic scattering mechanisms. Close inspection of the
temperature-dependent electron mobility between 50 and 375\,K allowed
us to distinguish dominant polar longitudinal optical phonon scattering
(PLOPS) from dominant ionized impurity scattering (IIS): Irrespective
of scattering mechanism, the conductivities in the $a$ and $b$ directions
agree within $2$\,\%. The ratio of the conductivities along the
$c$ and $b$ direction, however, is $0.96\pm0.01$ for PLOPS and
increases up to $1.12$ for IIS. The $a-b$ isotropy and $c-b$ anisotropy
of IIS may be related to the corresponding anisotropy of the dielectric
constant of $\beta$-Ga$_{2}$O$_{3}$.\cite{Ghosh2016,Fiedler2019}%
{} We experimentally demonstrate that significantly higher anisotropies
can be caused by extended structural defects in the form of low-angle
grain boundaries whose barriers were found to be multiple 10~meV
high.

Since the inherent electrical conductivity tensor of $\text{\ensuremath{\beta}-Ga}_{\text{2}}\text{O}_{\text{3}}$
can be considered fairly isotropic for practical applications no particular
transport direction is advantageous for increasing the electron mobility
for optimum performance of transport-based electronic devices, such
as transistors. %
Conductivity anisotropies of more than a few percent, as found in
Refs.~\cite{Ueda,Villora,Wong} are likely not intrinsic, but rather
related to experimental artifacts or to extrinsic causes, such as
extended structural defects like incoherent twin boundaries \cite{Fiedler}or
grain boundaries. The van der Pauw method used here is an experimentally
straightforward method to measure the in-plane transport anisotropy
of a given sample. We suggest to use the conductivity anisotropy of
a sample as a quality indicator to indicate extended defects oriented
along a certain crystallographic direction. 

Most of the conductivity anisotropies predicted by first principles
\cite{Kang,Ghosh2017} are significantly larger than those experimentally
found in this work. Their predicted dependence on electron concentration
\cite{Ghosh2017,Kang} requires further experimental investigation,
in particular towards electron concentrations above $10^{19}$\,cm$^{-3}$.
\begin{acknowledgments}
We thank F. Henschke for preparing the shadow mask, S. Rauwerdink,
W. Seidel, and W. Anders for lithography and contact deposition, A.
Riedel for wire bonding, as well as K. Irmscher for critically reading
the manuscript. This work was performed in the framework of GraFOx,
a Leibniz-ScienceCampus partially funded by the Leibniz association.
C.~G. gratefully acknowledges financial support by the Leibniz Association. 
\end{acknowledgments}

\bibliographystyle{apsrev}
\bibliography{GAO-conductivity-aniso-VdP_20190304}

\end{document}